\documentclass[%
 reprint,
 amsmath,amssymb,
 aps,
]{revtex4-2}

\usepackage{graphicx}% Include figure files
\usepackage{dcolumn}% Align table columns on decimal point
\usepackage{bm}% bold math
\begin{document}

\preprint{APS/123-QED}

\title{Self-regulating turbulence}% Force line breaks with \\
%\thanks{A footnote to the article title}%

\author{K. Steiros}
\email{k.steiros13@imperial.ac.uk}
% \altaffiliation[Also at ]{Aeronautics Department, Imperial College London}%Lines break automatically or can be forced with \\
 
\affiliation{%
Department of Aeronautics, Imperial College London, London SW7 2AZ, UK\\
}%

\date{\today}% It is always \today, today,
             %  but any date may be explicitly specified

\begin{abstract}
A landmark of out-of-equilibrium physics is Kolmogorov's phenomenological theory of turbulence. However, the past 20 years have provided evidence of a new, universal type of turbulence cascade, which does not abide to Kolmogorov physics. To address this issue, we revise the classical Kolmogorov cascade, by superimposing on it a mechanism of active information exchange between large and small scales. The new theory yields predictions for the dissipation rate, integral length scale and turbulence kinetic energy, as well as a criterion for the transition of the system from the new, to the classical physics in decaying turbulence. The assumptions and predictions are validated using large-scale simulations and data from the literature. 
\end{abstract}

%\keywords{Suggested keywords}%Use showkeys class option if keyword
                              %display desired
\maketitle

%\tableofcontents

\section{\label{sec:level1}Introduction}

Turbulence is connected to a flow mechanism which converts kinetic energy into heat, known as the turbulence cascade. Turbulent flows exhibit universal statistical properties; it can thus be expected that the turbulence cascade is a universal process as well. Understanding the cascade dynamics is a major challenge of out-of-equilibrium physics. 

According to the classical Richardson-Kolmogorov phenomenology \cite{kolmogorov1941dissipation}, the turbulence cascade is separated into collective modes. (i) Large scales which carry the bulk of the kinetic energy, (ii) small scales which dissipate the energy, and (iii) intermediate self-similar scales which mediate between the two. On average, energy moves from large to small scales, with energy dissipation being only a passive consequence of the energy injection into the cascade by the large scales. Thus, a one-way interaction between large and small scales is implied. The above description yields important predictions, such as a scaling law for the kinetic energy dissipation rate and the celebrated -5/3 law, both  validated by experiment in a wide variety of flows \cite{sreenivasan1984scaling,vassilicos2015dissipation,batchelor1953theory}.

Recent experimental and numerical results have revealed a new universal dissipation scaling, different to the classical one, appearing in extensive regions of decaying homogenous turbulence \cite{valente2012universal,isaza2014grid,goto2016unsteady}, boundary-free shear flows \cite{nedic2013axisymmetric,cafiero2020non}, as well as in forced periodic turbulence \cite{goto2015energy,goto2016local}. These results suggest the existence of a new type of cascade whose physics do not abide to the Richardson-Kolmogorov phenomenology.

Here, the above observations are explained theoretically, by revising the Richardson-Kolmogorov phenomenology, so as to include a feedback mechanism, enabling active interaction between large and small scales. The resulting framework yields the new dissipation scaling, as well as an equation for integral length scale evolution of the flow. Contrary to previous theories, turbulence invariants are not assumed. The decay of the turbulent kinetic energy is found to be governed by a generalized logistic equation, reflecting the self-regulation of the cascade.

\section{Self-regulation}
\label{sec:feedback}

The idea that a cascade feedback mechanism lies behind the new dissipation scaling has been anticipated by two existing non-Kolmogorov theories of turbulence, which have had some success in predicting the novel non-Kolmogorov dynamics, albeit both containing inconsistencies (see appendix \ref{app:1}).

%It is proposed that when the new dissipation scaling is valid, a feedback, linking large to small scales, is superimposed to Richardson-Kolmogorov physics. We first show that this idea has been anticipated by two existing non-Kolmogorov theories of turbulence, which have had some success in predicting the novel non-Kolmogorov dynamics, albeit both containing inconsistencies (see appendix \ref{app:1}).

George's theory \cite{george1992decay} (similar to the theory of Ref. \cite{barenblatt1974theory}) leads to the new dissipation scaling. The cascade in that case is assumed fully self similar, i.e. evolving as a ``coherent whole". In contrast, Richardson-Kolmogorov phenomenology presupposes independently evolving compartments (i.e. large/small scales), with a self-similar range of scales only in between. It might be thought that George's viewpoint implies a ``balanced" cascade of quasi-steady evolution. However, the novel dissipation scaling has been observed in cascades where intense fluctuations disturb the establishment of a balance, while the non-negligible cascade time-lag does not permit immediate relaxation \cite{goto2015energy,goto2016local}. As a result, evolution ``as a whole" suggests a regulatory mechanism of information exchange between large and small scales, which is thus implicit in George's theory.

%The theory of George [] anticipated the discovery of the new dissipation scaling. This theory assumes a fully self-similar cascade, i.e. evolving ``as a whole". In contrast, Richardson-Kolmogorov phenomenology presupposes independently evolving compartments (i.e. large/small scales), with a self-similar range of scales only in between. It might be thought that George's viewpoint implies a ``balanced" cascade whose evolution is quasi-steady. However, the novel dissipation scaling has been observed in cascades where intense fluctuations disturb the establishment of a balance [], while the non-negligible cascade time-lag does not permit immediate relaxation. As a result, evolution ``as a whole" suggests a regulatory mechanism of information exchange between large and small scales, which is thus implicit in George's theory.

Goto and Vassilicos \cite{goto2016unsteady} observed that large scales are not self-similar, and excluded them from George's analysis. Given the above discussion, this treatment removes George's implicit assumption of active communication between large and small scales, which, as proposed here, is the main cause of the new dissipation physics. Indeed, in order to predict the new scaling, Goto and Vassilicos \cite{goto2016unsteady} had to explicitly assume an \textit{ad-hoc} link between large and small scales (i.e. that their dissipation rates are proportional). We reiterate that such an explicit link was not necessary in George's theory, as full self-similarity already implied it, but became necessary as soon as full self-similarity was broken.

In appendix \ref{app:1} it is argued that, similar to the large scales, the small scales must also be removed from the self-similar analysis. We therefore return to our starting point, the Richardson-Kolmogorov picture of large and small scales, with a self-similar range only in between. However, our previous discussion suggests an important difference. The physics connected to the new dissipation scaling imply a feedback mechanism linking large and small scales, which must therefore be included in the analysis.

\section{Phenomenology and assumptions}
\label{sec:phenom}

\begin{figure}[b]
	\includegraphics[width=.78\columnwidth]{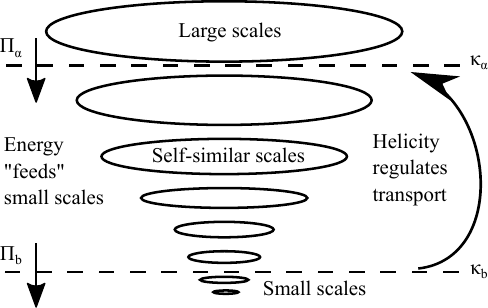}% Here is how to import EPS art
	\caption{\label{fig:Sketch} Proposed cascade picture. An intermediate range of self-similar scales is bounded by the non-dimensional wave numbers $\kappa_a$ and $\kappa_b$ from the large and small scales, respectively. A direct energy cascade ``feeds" the small scales, while an inverse helicity cascade regulates transport.}
\end{figure}

The ensuing analysis concerns homogeneous turbulence in cases where the new dissipation scaling has been observed. In decaying turbulence (grid and periodic box turbulence) this concerns the interval/region soon after turbulence starts to decay (i.e. close to the grid \cite{isaza2014grid}, or soon after the forcing stops \cite{goto2016unsteady}). At later times/distances the system transitions to the classical Kolmogorov dissipation scaling (note however that there are indications that even then Kolmogorov's assumptions are not fully valid \cite{goto2016unsteady}). In forced periodic box turbulence the flow-quantities undergo intense fluctuations \cite{goto2015energy,goto2016local} (even though the forcing is constant), during which the system always obeys the new dissipation scaling (see appendix \ref{app:2}). We note that the classical (Kolmogorov's) scaling is $\epsilon \sim K^{3/2}/L$, while the new dissipation scaling has been found from experiment to be $\epsilon \sim  \nu Re_{L0} K/L^2$ \cite{vassilicos2015dissipation}, where $K$ is the turbulent kinetic energy, $L$ the integral length scale, $Re_{L0}$ is the integral scale Reynolds number at the onset of decay and $\nu$ the kinematic viscosity.

In any case, homogeneous turbulence can be described by the scale-by-scale energy budget

\begin{equation}
	\frac{\partial K^>(k,t)}{\partial t} = \Pi(k,t) - \epsilon^>(k,t) \,.
	\label{eq:budget}
\end{equation}

\noindent With $E(k,t)$ the energy spectrum, $K^>(k,t) = \int_k ^ \infty E(k,t) dk$ and $\epsilon^>(k,t) = 2 \nu \int_k ^ \infty k^2 E(k,t) dk$ are the turbulent kinetic energy and dissipation rate, respectively, for wavenumbers larger than $k$. $\Pi(k,t)$ is the interscale flux of turbulent kinetic energy from wavenumbers smaller to wavenumbers larger than $k$. We omit ``per unit mass" throughout the text for brevity. Note that equation \ref{eq:budget} lacks a kinetic energy production term, i.e. this is assumed to act only in very small wavenumbers (large scales) not included in equation \ref{eq:budget}, or not be present at all, as in the case of purely decaying turbulence. 

We now perform a series of assumptions, which are validated using high Reynolds periodic box Direct Numerical Simulations (DNS) data of forced and decaying turbulence, (see \cite{goto2016unsteady} and appendix \ref{app:2} for more information on the numerical method and test cases). 

\paragraph*{Assumption 1:} Similar to the Richardson-Kolmogorov phenomenology, the cascade is separated into large scales, small scales and intermediate scales (see figure \ref{fig:Sketch}). The latter are assumed self-similar during decay, and bounded by the non dimensional wavenumbers $\kappa_a$ and $\kappa_b$, i.e.

$$
E(k,t) = A(t)f(kL,^*),  \hspace{0.4cm} \text{for $\kappa_a<kL<\kappa_b$} \,,
$$

\noindent where $\kappa = kL$ and $L(t) = \frac{3\pi}{4}\int_0 ^\infty k^{-1}E(k,t)dk/K(t)$. Following George \cite{george1992decay}, the argument $^*$ is included to indicate a dependency on initial conditions. 

Given the above assumption, we expect that the dissipation in the self-similar range $\epsilon^{ab}(t)$ will scale as the total dissipation of the cascade $\epsilon(t)$. That is, because the majority of $\epsilon^{ab}(t)$ is expected to occur at the largest wavenumbers of the self similar range, i.e. close to $\kappa_b$. The eddy turnover time at $\kappa \approx \kappa_b$ will thus regulate both $\epsilon^{ab}$ and $\Pi_b$, the latter being the interscale energy flux at $\kappa_b$ (see figure \ref{fig:Sketch}). We may thus expect $\epsilon^{ab} \sim \Pi_b$ (i.e. that their ratio is time-independent). Neglecting the dissipation of the large scales (i.e. for $\kappa<\kappa_a$) we have $\Pi_b \approx \epsilon - \epsilon^{ab}$ (Kolmogorov's small scale stationarity hypothesis \cite{kolmogorov1941degeneration}). Combination of the above yields $\epsilon^{ab} = \Phi \epsilon$, where $\Phi$ is a constant of proportionality.

\begin{figure*}
	\centerline{
		\begin{tabular}{ll}
			$\qquad$ (a) & $\qquad$ (b) \\
			\includegraphics[width=.96\columnwidth]{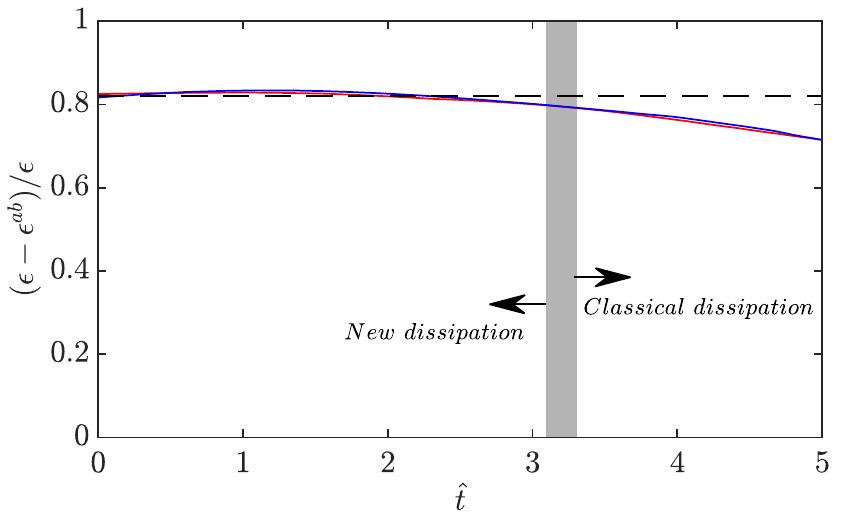} &
			\includegraphics[width=.96\columnwidth]{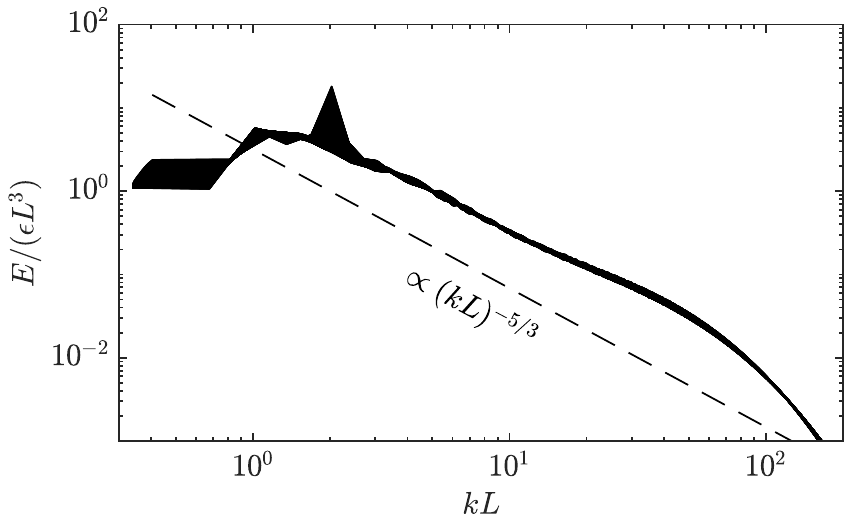}
		\end{tabular}
	}
	\caption{(a) Normalized dissipation against number of turnover times, for periodic box decaying turbulence, with simulation sizes $N=2048$ (red line) and $N=1024$ (blue line). The grey stripe marks the transition region from the new dissipation scaling to the classical one. (b) Normalized energy spectra for many instances while the new dissipation scaling is valid. The novel scaling ceases to be valid approximately at the same time when the spectral peak at $kL\approx2$ disappears.}
	\label{fig:A1}
\end{figure*}

This result is validated in figure \ref{fig:A1}a where the appropriately normalized dissipation of periodic-box decaying turbulence is plotted as a function of the number of eddy turnover times $\hat{t} = \int _{0} ^t \frac{u}{L}dt$, where $\frac{3}{2}u^2 = K$. Two simulation sizes $N^3$ are plotted, i.e. $N=2048$ and $N = 1024$ (larger size corresponds to larger Reynolds number). The cutoff non-dimensional wavenumber $\kappa_b$ naturally increases with initial Reynolds number, and is taken to be equal to to 41 and 22, for the high and low Re cases, respectively (i.e. approximately at the wavenumber where the -5/3 spectral scaling starts to break down, see figure \ref{fig:A1}b). For both cases the normalized dissipation is relatively constant while the new dissipation scaling holds, giving some support to assumption 1. 

However, for larger times $\epsilon^{ab} \sim \epsilon$ ceases to be valid, and this coincides with the shift of the system to the classical (Kolmogorov) dissipation scaling. The reason for this is that assumption 1 treats $\kappa_a$ and $\kappa_b$ as time-independent. When the new dissipation scaling is valid, this is indeed true. In that case, the beginning of the self-similar range (and thus $\kappa_a$) occurs shortly after the spectral peak (see figure \ref{fig:A1}b). Our DNS results show that while the latter diminishes with time, it always stays centred around the same normalized wavenumber $kL$. At the same time we expect $\kappa_b$ to be roughly proportional to $L/\lambda$, where $\lambda$ is the Taylor microscale. In section \ref{sec:diss} it is indeed shown that $L/\lambda$ stays constant with time when the new dissipation scaling is valid. The transition of the system to the classical scaling coincides with the disappearance of the spectral peak, and the start of a decreasing trend of $L/\lambda$ with time: $\kappa_a$ and $\kappa_b$ are thus no longer time-independent and assumption 1 is invalid.

Given the above analysis, we may obtain a scaling law for the energy spectrum $E(k,t)$. Using assumption 1 the dissipation of the self similar part of the cascade is given as

$$
\epsilon^{ab} = 2 \nu L^3 A \int_{\kappa_a} ^{\kappa_b} \kappa^2 f(\kappa,^*)d\kappa \,,
$$

\noindent which yields an expression for the time-evolution parameter of the spectrum $A(t)$. However, we have just shown that when the new dissipation scaling is valid we have $\epsilon^{ab} = \Phi \epsilon$, and thus we obtain

\begin{equation}
	E(k,t) = \frac{\Phi \epsilon L^3}{2 \nu I_2}f(\kappa,^*),  \hspace{0.4cm} \text{for $\kappa_a<kL<\kappa_b$} \,,
	\label{eq:A1sb}
\end{equation}

\noindent where $I_2=\int_{\kappa_a} ^{\kappa_b} \kappa^2 f(\kappa,^*)d\kappa$. This scaling was introduced first in Ref. \cite{george1992decay}, using qualitatively similar arguments. Goto and Vassilicos \cite{goto2016unsteady} provided evidence for equation \ref{eq:A1sb}, for high enough wavenumbers, and when the new dissipation is valid. In figure \ref{fig:A1} we reproduce their data (decaying periodic box turbulence at $N=2048$) for completeness. The DNS data offer acceptable support for this scaling. Note that these spectra are the the hardest to collapse, as in this time-interval the Reynolds number varies the most during decay.

\begin{figure*}
	\centerline{
		\begin{tabular}{ll}
			$\qquad$ (a) & $\qquad$ (b) \\
			\includegraphics[width=.96\columnwidth]{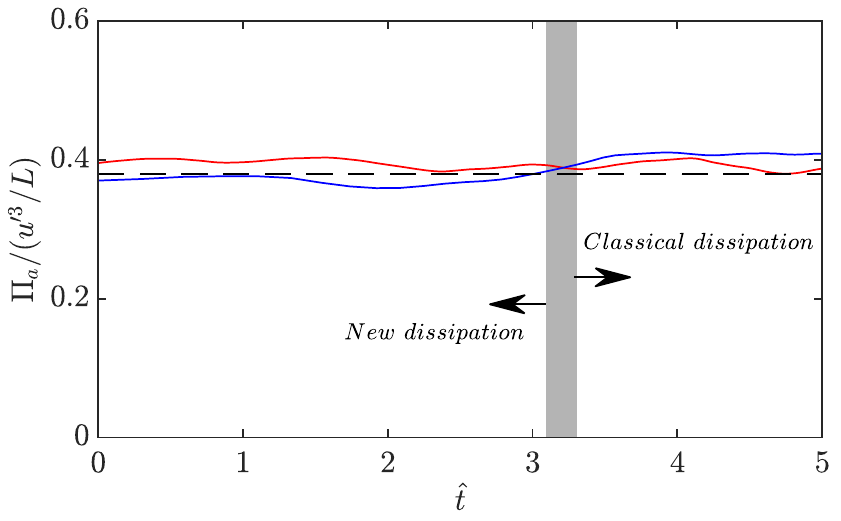} &
			\includegraphics[width=.96\columnwidth]{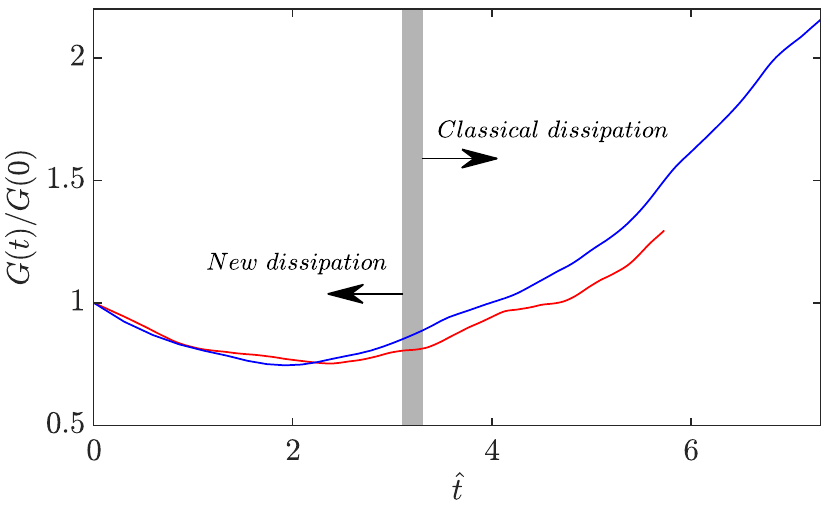}
		\end{tabular}
	}
	\caption{ (a) Normalized interscale energy flux of the large scales $\Pi_a$ and (b) normalized parameter $G(t)$ for decaying periodic turbulence of domain size $N=2048$ (red) and $N=1024$ (blue) (the forcing stops at $\hat{t}=0$). The dashed line in (a) corresponds to an ordinate value of 0.38.}
	\label{fig:A2}
\end{figure*}

\paragraph*{Assumption 2:} Much similarly to the Richardson-Kolmogorov phenomenology, it is assumed that a wavenumber $\kappa_a$ exists in the upper part of the self-similar range, such that 

$$
\Pi_a = C_x u^3/L \,,
$$

\noindent where $\Pi_a$ is evaluated at $\kappa_a$, and $C_x$ is a coefficient of proportionality. While this expression is generally accepted for Kolmogorov turbulence \cite{vassilicos2015dissipation,pope2001turbulent}, it is not straightforward that it holds when the new dissipation scaling is valid. For instance, Goto and Vassilicos \cite{goto2016unsteady} have shown that in decaying turbulence the above does not hold for a wide wavenumber range in the self-similar part of the cascade. However, figure \ref{fig:A2}a shows that the above relationship holds for decaying periodic-box turbulence if $\kappa_a$ is taken shortly after the spectral peak. Specifically, here $\Pi_a$ is calculated for $\kappa_a\approx 3.3$, with the spectral peak being centred around $kL = 2$. A similar result can be also obtained for forced turbulence. There, Goto and Vassilicos \cite{goto2016local} have shown that assumption 2 is always valid, when calculated at an appropriate wavenumber. The coefficient of proportionality in forced turbulence was found to be very close to the one calculated here ($C_x \approx 0.38$).

%Goto and Vassilicos \cite{goto2016local} showed that in forced turbulence (where the new dissipation scaling always holds) $\Pi_a = C_x u'^3/L$ with $C_x=0.38$. Figure \ref{fig:A2}a shows for the first time that this is also true for unforced, decaying periodic box turbulence, for the two domain sizes tested, i.e. $N=2048$ and $N=1024$. $\Pi_a$ is calculated for $\kappa_a=3.5$ and  $\kappa_a=3$, respectively, i.e. for wavenumbers slightly larger than the spectral peak in figure \ref{fig:A1}b. The spectral peak does not collapse with the scaling of equation \ref{eq:A1sb} and thus cannot be included in the self-similar range. The coefficient of proportionality is found to be close to $C_x=0.38$.

\paragraph*{Assumption 3:} When the flow exhibits the new dissipation scaling, the large scale interscale flux, $\Pi_a$, and the dissipation rate, $\epsilon$, are connected via the expression

$$
\Pi_a \sim \epsilon Re_L \,.
$$

\noindent This is the essential point of departure from the Kolmogorov phenomenology, which simply assumes $\Pi_a \sim \epsilon$. Assumption 3 is admittedly \textit{ad-hoc}; it is necessary for the analysis to yield the new dissipation scaling (see section \ref{sec:diss}). Conversely, assumption 2 transforms the new dissipation scaling into a simpler statement (assumption 3) which is much easier to interpret physically.

In figure \ref{fig:A2}b we validate the above assumption by plotting the normalized flux $G(t) = \Pi_a/(\epsilon Re_L)$ for decaying periodic box turbulence (as above, $\kappa_a=3$ and  $\kappa_a=3.5$ for the two domain sizes). The normalized flux drops slightly and then remains relatively constant, as long as the system is characterized by the new dissipation scaling, providing some backing to assumption 3 (we note that for slightly larger $\kappa_a$ the constancy of $G(t)$ improves).

We now argue that assumption 3 is the expression of a negative feedback in the cascade. This is more evident in forced turbulence conditions where the turbulence parameters exhibit quasiperiodical oscillations, even if the forcing remains invariant in time (see \cite{goto2016local} and appendix \ref{app:2}). This behaviour is reminiscent of predator-prey systems \cite{brauer2012mathematical} where a negative feedback works to establish ``balance" in the system and oscillations are observed. Assumption 3 expresses a negative feedback according to the following causal chain. In forced turbulence, if the interscale flux $\Pi_a$ were to increase, then this would cause an increase in $\epsilon$ (after a time-lag). Turbulence would thus start to decay, causing a drop in $Re_L$ (in appendix \ref{app:2} we show that $\epsilon$ and $Re_L$ are indeed somewhat anticorrelated in forced turbulence). Assumption 3 would then halt the increase of $\Pi_a$, moving the system towards its previous state (negative feedback). The opposite would occur if $\Pi_a$ were to decrease.

The above causal chain requires a physical mechanism which would permit an information exchange between large and small scales. We now postulate such a mechanism based on helicity, the latter being the inner product of velocity and vorticity, $H = \boldsymbol{u}\boldsymbol{\omega}$. High values of $H$ deplete the nonlinearity of Navier-Stokes equations, suppressing the interscale transfer of the cascade \cite{moffatt2014helicity}. Small-scale helicity thus offers a pathway for active communication between large and small scales.

We first discuss the results of two recent works which, when combined, indicate this role of small scale helicity in the cascade. First, the DNS of \cite{portela2018turbulence} imply that, when the new dissipation scaling holds, small scale structures of high helicity exist in the flow, whose appearance is correlated to that of large coherent vortices in the flow. It is interesting that the current DNS results actually show that the new dissipation scaling holds for as long as the vortex peak of figure \ref{fig:A1}b (footprint of large coherent vortices) appears in the spectrum. As soon as the peak disappears, the system transitions to the classical dissipation scaling. Second, the analysis of \cite{bos2017dissipation} (see also \cite{yoshizawa1994nonequilibrium}) links the new dissipation scaling to a -7/3 slope in the energy spectrum, coexisting with the -5/3 slope, and therefore masked by it. The earlier work of \cite{brissaud1973helicity} actually suggests that a -7/3 slope is the footprint of an inverse helicity cascade, i.e. helicity transport from small to large scales. 

Combining the above points, we may postulate the following feedback mechanism, also depicted in figure \ref{fig:Sketch}. An instability mechanism causes the large scales to create small helical structures of high helicity. Helicity then cascades up towards the large scales, finally intercepting the interscale flux $\Pi_a$. Assumption 3 (and thus the new dissipation scaling) could be thought to be the expression of these dynamics, in the sense that $\Pi_a$ is larger when dissipation is high (so that the small helical structures are destroyed) and when $Re_L$ is large (so that scale separation and thus the inverse cascade lag is large). Validation of this physical mechanism is left as a task for future research.

\section{Results}

\subsection{Dissipation rate} 
\label{sec:diss}

First, we consider forced turbulence. Assumption 3 is 

$$
\epsilon = C \frac{\Pi_a}{Re_L}\,.
$$

\noindent Considering a time-averaged cascade where large scale dissipation is negligible, we have $\overline{ \epsilon } = \overline{\Pi}_a $, where the bar denotes the time-averaging operation. We expect that the cascade time-lag breaks any correlation between $\Pi_a$ and $Re_L$ in forced turbulence (see appendix \ref{app:2} for validation of this assumption). Thus, time averaging of the above expression yields $C = 1/\overline{Re^{-1}_L}$. This is approximately  $C \approx \overline{Re}_L$ (the forced turbulence data of \cite{goto2016local} confirm this simplification). Consequently, combination of assumptions 2 and 3 yields

\begin{equation}
	\epsilon \sim \overline{ uL}  \frac{u^2}{L^2} \,,
	\label{eq:diss1}
\end{equation}

\noindent which is the new dissipation scaling. For decaying turbulence, we achieve a similar result if, instead of time averaging, we perform ensemble averaging at time $t=0$, where the turbulence is still forced. Thus, we have $\langle  \epsilon_0 \rangle = \langle \Pi_{a0} \rangle $, where the subscript 0 signifies the time $t=0$, and we obtain 

\begin{equation}
	\epsilon \sim u_0 L_0   \frac{u^2}{L^2} \,.
	\label{eq:diss2}
\end{equation}

\begin{figure*}
	\centerline{
		\begin{tabular}{lll}
			$\qquad$ (a) & $\qquad$ (b) & $\qquad$ (c) \\
			\includegraphics[width=.64\columnwidth]{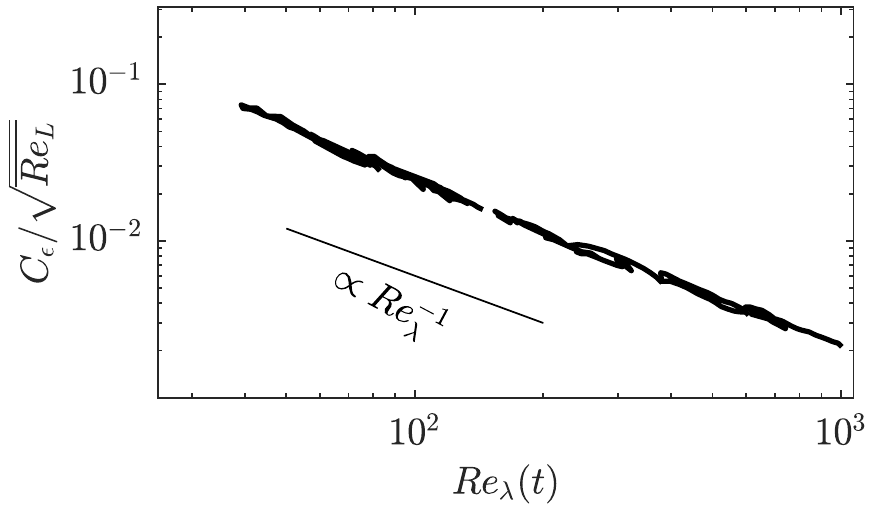} &
			\includegraphics[width=.64\columnwidth]{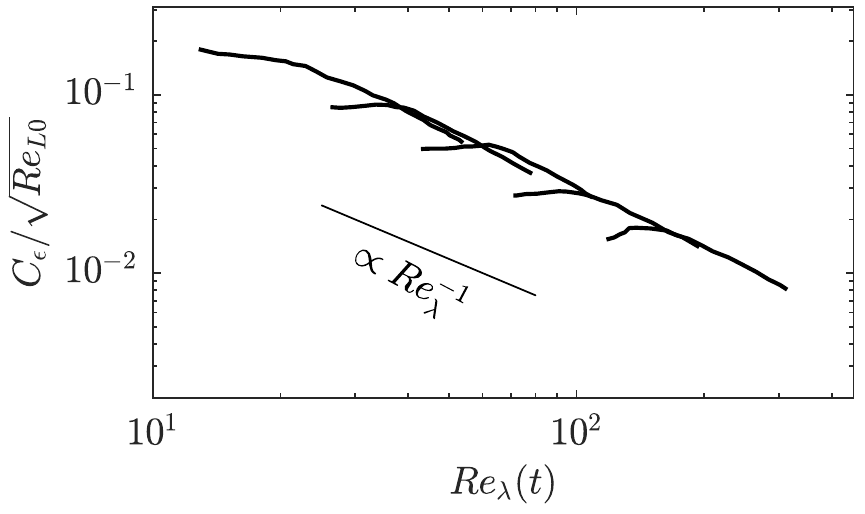}&
			\includegraphics[width=.64\columnwidth]{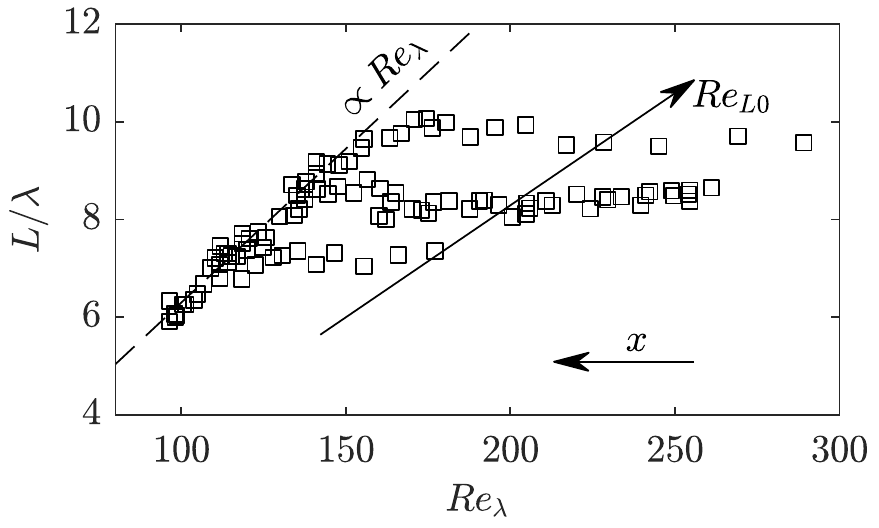}
		\end{tabular}
	}
	\caption{Time evolution of the normalized $C_\epsilon$ for (a) forced periodic and (b) decaying periodic turbulence simulations of various initial Reynolds numbers (from \cite{goto2016local} and \cite{goto2016unsteady}). (c) Spatial evolution of $L/\lambda$ for various grids in grid-generated turbulence experiments (from \cite{valente2012universal}).}
	\label{fig:V1}
\end{figure*}

\noindent In turbulence literature, dissipation scalings are commonly expressed using the dissipation coefficient $C_\epsilon = \epsilon L/u^3$, which is constant in Kolmogorov turbulence. Using the definition of the Taylor length scale $\lambda^2  \equiv 15 \nu u^2/\epsilon$, we obtain

\begin{equation}
	L/\lambda \sim C_\epsilon Re_\lambda \,,
	\label{eq:diss3}
\end{equation}

\noindent which shows that $L/\lambda$ increases linearly with $Re_\lambda$ in Kolmogorov turbulence. On the other hand, when the new dissipation scaling holds (i.e. equations \ref{eq:diss1} and \ref{eq:diss2}), we have

\begin{equation}
	C_\epsilon \sim \sqrt{Re_{L0}  }  Re_\lambda ^{-1}\,,
	\label{eq:diss4}
\end{equation}

\noindent where $Re_{L0}$ may denote either the time-averaged Reynolds number for forced turbulence, or the initial condition Reynolds number, for decaying turbulence. Substitution to expression \ref{eq:diss3} shows that $L/\lambda$ is constant during decay when the new dissipation scaling holds. In figure \ref{fig:V1}, we validate the above predictions using data from the literature for forced periodic, decaying periodic, and grid turbulence (see appendix \ref{app:2} and \cite{goto2016unsteady} for more info on the data-sets used). For forced turbulence (figure \ref{fig:V1}a) the different simulation runs are always characterized by the new dissipation scaling (equation \ref{eq:diss4}). In decaying turbulence (figure \ref{fig:V1}b) all five simulations begin with the new dissipation scaling, and later transition to the Kolmogorov scaling ($C_\epsilon \approx const$). As mentioned in the previous section, this state change coincides with the disappearance of the coherent vortices from the flow. In grid turbulence (figure \ref{fig:V1}c), for all tested grids the flow begins with the new dissipation scaling ($L/\lambda = const$) and at larger distances from the grid it transitions to the Kolmogorov scaling ($L/\lambda \sim  Re_\lambda$).

\subsection{Integral length scale}
\label{sec:integral}

The two dissipation scalings (classical, new) discussed in the previous sections, provide a starting point for the prediction of the kinetic energy evolution of homogenous decaying turbulence, in the sense that $dK/dt = - \epsilon$. However, this equation cannot be integrated, given that $\epsilon$ is a function of $L$, which is itself an unknown function of time. This closure problem has been conventionally resolved via the \textit{ad hoc} assumption of ``turbulence invariants" \cite{sinhuber2015decay,saffman1967large}. This assumption is often arbitrary, given that an infinite number of invariants exist in turbulent flows \cite{vassilicos2011infinity}. In contrast to previous theories, the current framework yields a prediction of $L$ implicitly and does not rely on the assumption of invariants.

Neglecting the kinetic energy of the small scales ($kL>\kappa_b$), we obtain an estimate for the turbulence kinetic energy for scales larger than $k$, by integrating equation \ref{eq:A1sb} from $k$ to $\kappa_b/L$, i.e.

\begin{equation}
	K^>(k,t) \approx \frac{\Phi \epsilon L^2}{2 \nu I_2} I_0(kL) \,,
	\label{eq:L1} 
\end{equation}

\noindent where $I_0(kL) = \int_{\kappa} ^{\kappa_b} f(\kappa,^*) d\kappa$. Injection of $\partial K^>/\partial t$ evaluated at $\kappa_a$ (we remind that assumption 1 states that both $\kappa_a$ and $\kappa_b$ are time-independent), along with the new dissipation scaling ($\epsilon = u_0 L_0 C_x \frac{u^2}{L^2}$) and assumption 3 ($\Pi_a = \frac{uL}{u_0 L_0} \epsilon$) to the scale-by-scale energy budget (equation \ref{eq:budget}) yields

%$$
%\frac{\partial K^>}{\partial t} = -\epsilon \Phi \left[ \frac{C_x u_0 L_0 I_0(kL)}{3 \nu I_2(2)} + \frac{kL  f(kL)}{4 \nu I_2(2)}  \frac{dL^2}{dt} \right]\,.
%$$

\begin{equation}
	\frac{1}{\nu} \frac{dL^2}{dt} = A - B Re_\lambda \,,
	\label{eq:L2}
\end{equation}

\noindent where $A = 4 \frac{ I_2 \frac{1}{\Phi} - \frac{1}{3}C_x Re_{L0} I_0   }{\kappa_a f(\kappa_a,^*)}$ and $B = \frac{4  I_2 }{\Phi \kappa_a f(\kappa_a,^*)}\sqrt{\frac{C_x}{15Re_{L0}}}$ are positive constants dependent on initial conditions. In the above, $I_0 = \int_{\kappa_a} ^{\kappa_b} f(\kappa,^*) d\kappa$.

The above analysis can also yield a prediction for the point of transition from the new to the classical dissipation scaling. Equation \ref{eq:L1} relies on the assumption $\epsilon^{ab} \sim \epsilon $ (see section \ref{sec:phenom}) which does not hold in the classical dissipation scaling (see figure \ref{fig:A1}). However, we may consider $\epsilon^{ab} \sim \epsilon $ to be approximately valid for a small time interval after the state change. We may thus repeat the analysis of this section, but using the ``classical" expressions for the dissipation and interscale transfer, i.e. $\epsilon \sim u^3/L$ and $\Pi_a \sim \epsilon$. The result (see appendix \ref{app:3}) is 

\begin{equation}
	\frac{1}{\nu} \frac{dL^2}{dt} = -A' + B' Re^2 _\lambda \,,
	\label{eq:L3}
\end{equation}

\noindent with $B'$ a positive constant for sufficiently high Reynolds numbers. We thus conclude that the transition from the new to the classical dissipation scaling occurs when the slope of $\frac{dL^2}{dt}$ changes sign. This is in agreement with the observation of \cite{goto2016unsteady}, that the state change coincides with the location where $\frac{dL^2}{dt}$ assumes its maximum value. We emphasize that equation \ref{eq:L3} is not valid, in general, during the classical decay, but only for a very small interval after the state change of the system.

\begin{figure*}
	\centerline{
		\begin{tabular}{ll}
			$\qquad$ (a) & $\qquad$ (b) \\
			\includegraphics[width=.96\columnwidth]{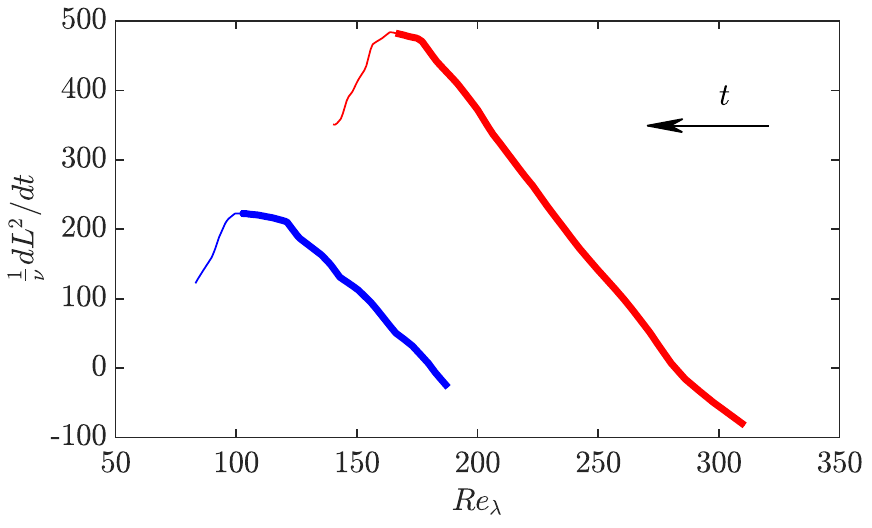} &
			\includegraphics[width=.96\columnwidth]{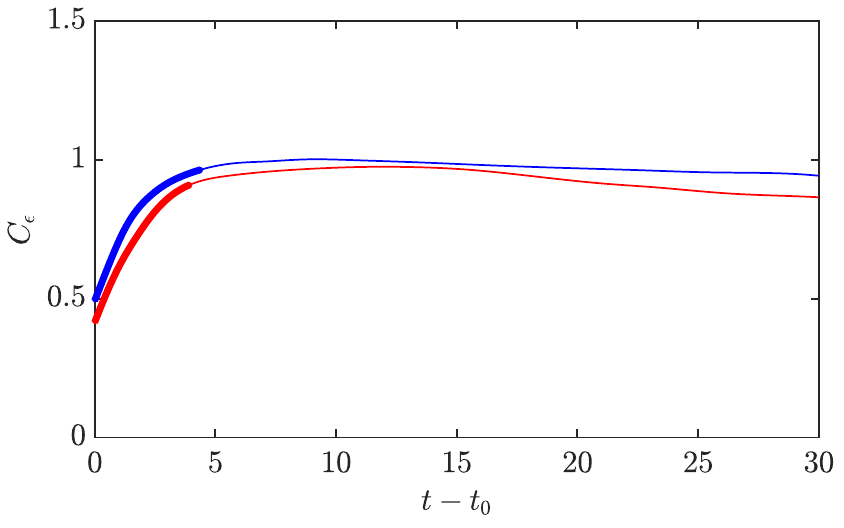}
		\end{tabular}
	}
	\caption{(a)  $\frac{1}{\nu}\frac{dL^2}{dt}$ and (b) $C_\epsilon$ for domain sizes $N=2048$ (red) and $N=1024$ (blue) (the forcing stops at $t=t_0$). The thick part of the lines marks the range where $\frac{dL^2}{dt}$ grows.}
	\label{fig:L1}
\end{figure*}

The above predictions are validated in figure \ref{fig:L1}a using the two decaying periodic box data-sets. In accordance with equation \ref{eq:L2}, $\frac{dL^2}{dt}$ is a linear decreasing function of $Re_\lambda$, for as long as the new dissipation scaling holds (see figure \ref{fig:L1}b). When the system transitions to the classical scaling (i.e. $C_\epsilon = cont$), $\frac{dL^2}{dt}$ becomes an increasing function of $Re_\lambda$, in agreement with equation \ref{eq:L3}. The maximum value of $\frac{dL^2}{dt}$ marks the state change.

\subsection{Turbulent kinetic energy}
\label{sec:Equation}

Having validated the scalings \ref{eq:diss2} and \ref{eq:L2}, we may combine them to obtain an expression for the evolution of the turbulent kinetic energy during decay. Elimination of time yields (equations \ref{eq:diss3} and \ref{eq:diss4} are also used)

\begin{equation}
	\frac{du^2}{dL^2} = \frac{-u^2}{C_1L^2 - C_2 u L^3}\,,
	\label{eq:K1}
\end{equation}

\noindent where $C_1 =  \frac{6I_2/(\Phi Re_{L0}C_x) - 2I_0}{\kappa_a f(\kappa_a,^*)}$ and $C_2 = \frac{6I_2}{\Phi \kappa_a f(\kappa_a,^*) Re_{L0} C_x u_0 L_0}$. In the above we have considered $\frac{3}{2} \frac{du^2}{dt} = - \epsilon$, i.e. decaying turbulence without turbulence production. It can be checked by substitution that a solution to the above equation is 

\begin{equation}
	\frac{C_1-1}{uL} = C_2  - \left(\frac{u}{C}\right)^{C_1-1} \,,
	\label{eq:K2}
\end{equation}

\noindent with $C$ a positive constant of integration. Evidently, the current framework correctly predicts a continuously decreasing Reynolds number during decay, in contrast to previous theories for the new dissipation scaling (see appendix \ref{app:1}). Combination of equations \ref{eq:K1} and \ref{eq:K2} yields

\begin{equation}
	\frac{du^2}{dt} \sim  - u^4 \left[1- \left(\frac{u}{c}\right)^{C_1-1} \right]^2 \,,
	\label{eq:K3}
\end{equation}

\noindent with $c$ a positive constant. Expression \ref{eq:K3} is a generalized logistic equation \cite{tsoularis2002analysis} (if $C_1=1$ it reduces to a generalized Gompertz equation), and it expresses the regulation introduced by assumption 3 via the term $\left[1- \left(\frac{u}{c}\right)^{C_1-1} \right]^2$. For this term (and thus for regulation) to be negligible, the second term on the right hand side of equation \ref{eq:K2} would also need to be negligible. Thus, $Re_L$ would have to remain approximately constant during decay. Then, assumption 3 would reduce to $\Pi_a \sim \epsilon$ (i.e. Kolmogorov turbulence) and thus the regulation that it otherwise expresses (see discussion in section \ref{sec:phenom}) would be lost. 

We might inquire how does self-regulation affect the distribution of kinetic energy across the scales. Combination of equations \ref{eq:diss2} and \ref{eq:L1} yields 

\begin{equation}
	\frac{K^{ab}}{K} = \frac{\Phi Re_{L0} C_x I_0}{3 I_2} = const \,,
	\label{eq:K4}
\end{equation}

\noindent where $K^{ab}$ is the energy of the self-similar scales. Thus, we obtain the result that, despite a qualitative change in the spectrum (i.e. disappearance of the spectral peak, see figure \ref{fig:A1}b), self-regulation in the cascade guarantees that the ratio of the kinetic energy between large and self-similar scales be constant.  

\begin{figure*}
	\centerline{
		\begin{tabular}{ll}
			$\qquad$ (a) & $\qquad$ (b) \\
			\includegraphics[width=.96\columnwidth]{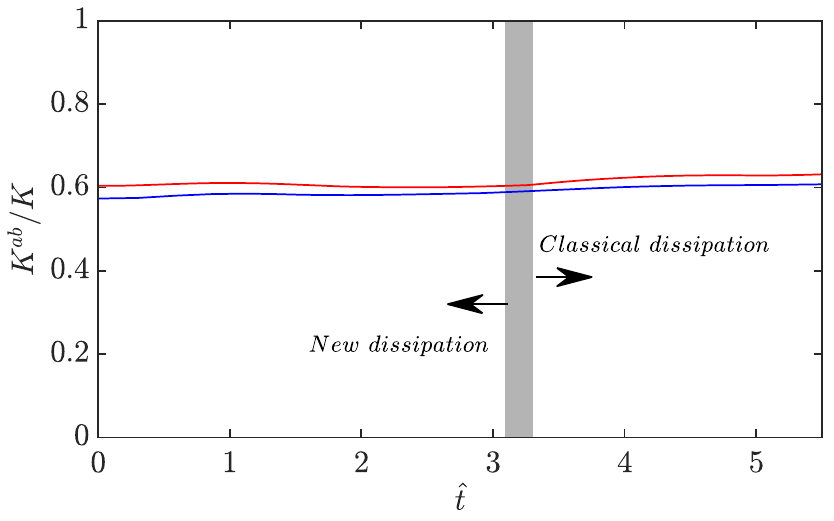} &
			\includegraphics[width=.96\columnwidth]{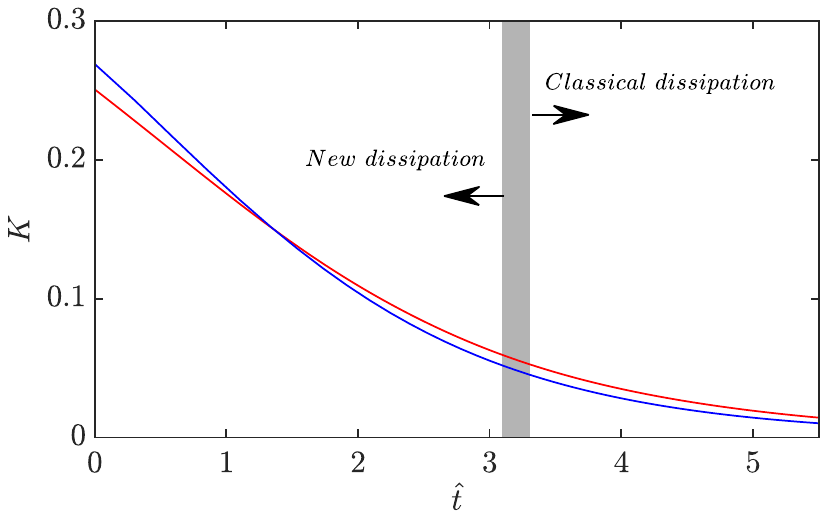}
		\end{tabular}
	}
	\caption{(a) Ratio of the kinetic energy of the self-similar range over the total cascade kinetic energy and (b) total cascade kinetic energy, versus number of turnover times for domain sizes $N=2048$ (red) and $N=1024$ (blue). }
	\label{fig:K}
\end{figure*}

In figure \ref{fig:K}a we show that the ratio $K^{ab}/K$ indeed stays relatively constant when the separation wavenumber is taken immediately after the spectral peak (see figure \ref{fig:A1}b), i.e. at $\kappa_a = 2.3$, for both of our decaying-turbulence data sets. Note that while the new dissipation scaling holds, the cascade undergoes the most change during decay, losing roughly 80\% of its initial kinetic energy (see figure \ref{fig:K}b).

\section{Concluding discussion}

By superimposing a feedback mechanism on the classical Richardson-Kolmogorov phenomenology we derived expressions for various flow quantities (dissipation rate, integral length scale, kinetic energy) of the non-Kolmogorov universal cascade that has been recently discovered \cite{seoud2007dissipation}. We reiterate that the new type of cascade regulates forced turbulence \cite{goto2015energy}, the region of decaying turbulence where the bulk of kinetic energy is lost \cite{isaza2014grid}, and almost the whole extent of turbulent wakes \cite{redford2012universality,dairay2015non}. Therefore, it might be considered more relevant for engineering applications (and thus turbulence modelling) than classical Kolmogorov turbulence.

In the special case of forced turbulence, the current cascade picture resembles low-order predator-prey dynamics; prey (large scales) feeds the predator (small scales) in a self-regulating manner. These dynamics would explain the quasi-periodic oscillations of the turbulence quantities observed (see \cite{goto2015energy,goto2016local} and appendix \ref{app:2}), which indeed resemble, qualitatively, the response of predator-prey systems \cite{brauer2012mathematical}. The present analysis describes how small scales remove energy from the system during this regime. The question of how large scales replenish energy when forcing is present, remains open and will comprise the topic of future research. Another topic which remains open is the exact instability mechanism which generates the feedback, here attributed to an inverse helicity cascade.

%\begin{acknowledgments}
%I am grateful to Prof Christos Vassilicos for introducing me to the problem, for helpful comments and discussions, and for reading an early version of the manuscript. I would also like to thank Prof Susumu Goto for sharing the DNS datasets that were used in the article.
%\end{acknowledgments}

\appendix
\section{Over-constraint of self-similarity}
\label{app:1}

We consider two previous non-Kolmogorov theories of turbulence which predict the new dissipation scaling and we investigate certain assumptions which lead to inconsistencies. Goto and Vassilicos \cite{goto2016unsteady}, considered a cascade where all non-dimensional wavenumbers larger than a given small wavenumber, and up to infinity, are self-similar. Then, for that range of scales, which includes the dissipative range, all length scales are proportional to the integral length scale $L$. Consequently,

$$
L/\eta = const \,,
$$

\noindent where $\eta = (\frac{\nu^3}{\epsilon})^{1/4}$ is the Kolmogorov scale, characteristic of the dissipative range. The new dissipation scaling is $\epsilon \sim \nu Re_{L0} \frac{K}{L^2}$ (see section \ref{sec:diss}). Combining this with the definition of the Kolmogorov scale yields

$$
L/\eta \sim \sqrt{Re_L} \,,
$$

\noindent where $Re_L$ is the Reynolds number based on the integral length scale and turbulence kinetic energy. Therefore, inclusion of the small scales in the self-similar range overconstrains the system and implies constancy of Reynolds number during decay. This is not in agreement with observations. However, we do note that because the kinetic energy of the small scales is small, their inclusion, or not, in the self-similar range does not drastically alter the other predictions of this theory. Indeed, the bulk of the predictions of Goto and Vassilicos \cite{goto2016unsteady} agree, in general, with observations, despite this inconsistency.

It is instructive to to investigate the conclusions of the current theory if constant Reynolds number were to be imposed to it. Equation \ref{eq:L2} then would yield $dL^2/dt = const$ during decay. Again, this is not in agreement with observations (see figure \ref{fig:L1}a). We note that this expression for the integral length scale is one of the main results of the fully self-similar theory of George \cite{george1992decay}. Equations 6 and 13 of Goto and Vassilicos show that this erroneous result can be traced to George's inclusion of large scales in the self-similar analysis. 

The above discussion suggests that the cascade becomes overconstrained if either the large or the small scales (or both) are included in the self-similar analysis. This leads to the particular (unrealistic) decay where Reynolds number remains constant. The current framework relaxes self-similarity, and assumes it valid only at an intermediate range of scales, much like Kolmogorov \cite{kolmogorov1941dissipation}. The predictions of the previous theories are then recovered only if Reynolds number is explicitly assumed constant during decay. 

\section{Validation data-sets}
\label{app:2}

\begin{figure*}
	\centerline{
		\begin{tabular}{ll}
			$\qquad$ (a) & $\qquad$ (b) \\
			\includegraphics[width=.96\columnwidth]{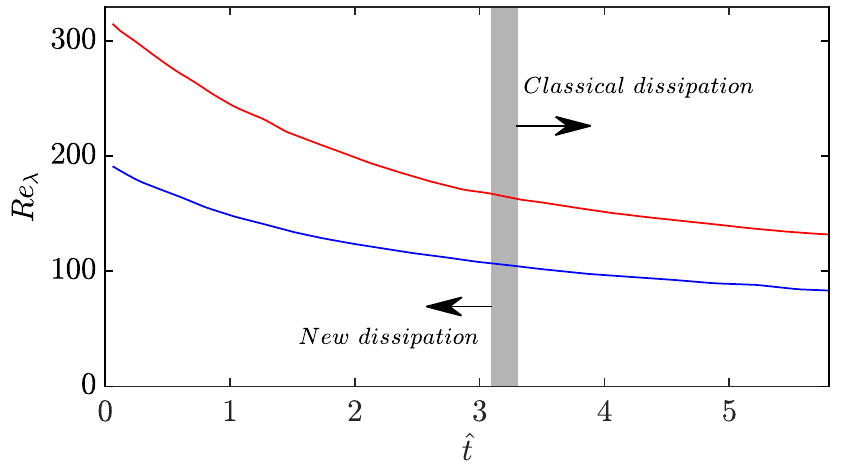} &
			\includegraphics[width=.96\columnwidth]{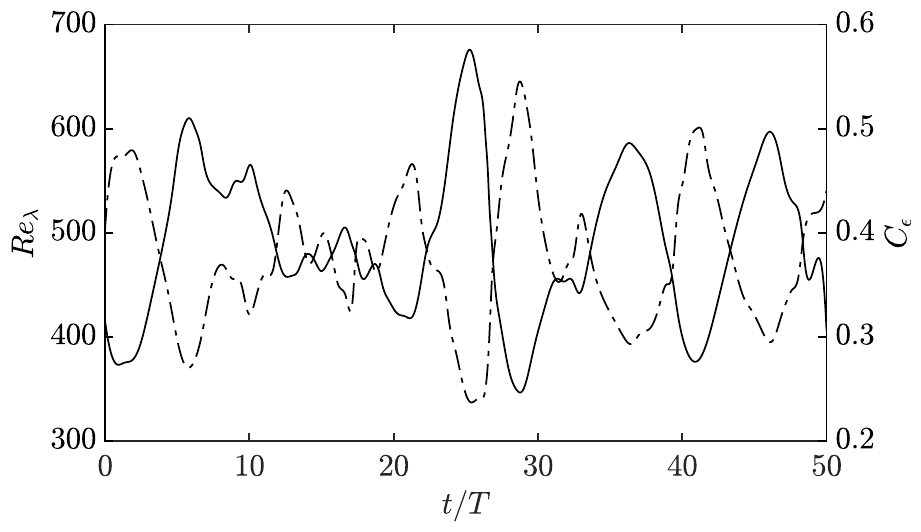}
		\end{tabular}
	}
	\caption{(a) Decay of $Re_\lambda$, versus number of turnover times, for decaying periodic box turbulence of domain size $N=2048$ (red) and $N=1024$ (blue) (from \cite{goto2016unsteady}). (b) Evolution of $Re_\lambda$ (solid line) and $C_\epsilon$ (dashed-dotted line) for periodic box turbulence of constant forcing, versus time normalized with the mean turnover time (from \cite{goto2016local}).}
	\label{fig:app2}
\end{figure*}

For validation purposes, two data-sets of periodic-box decaying turbulence are used, the details of which are presented in \cite{goto2016unsteady}. For both cases, a forcing $f = (-\sin(k_fx) \cos(k_fy), \cos(k_fx)\sin(k_fy),0)$ with $k_f=4$ is imposed on the Navier-Stokes equations, and is turned off at $t=t_0$, allowing the turbulence to decay. The first data set concerns an ensemble of ten simulations of $N^3=1024^3$; the presented results are ensemble averages. The second data-set concerns a simulation size of $N^3=2048^3$ that contains a single run. The larger simulation size corresponds to a larger Reynolds number. The spatial resolution $k_{max}\eta$ is slightly larger than one at $t_0$, while $k_{max}\eta$ increases during decay. The decay of $Re_\lambda$ for the two data sets is depicted in figure \ref{fig:app2}a.

\begin{figure*}
	\centerline{
		\begin{tabular}{ll}
			$\qquad$ (a) & $\qquad$ (b) \\
			\includegraphics[width=.96\columnwidth]{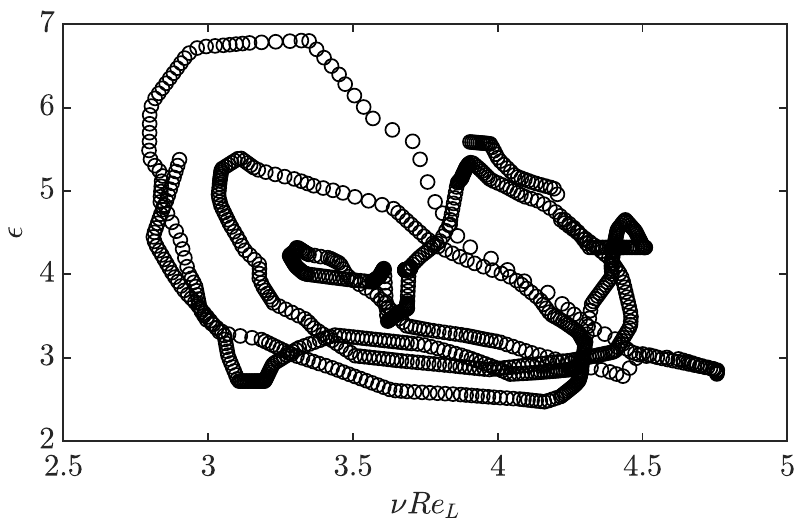} &
			\includegraphics[width=.96\columnwidth]{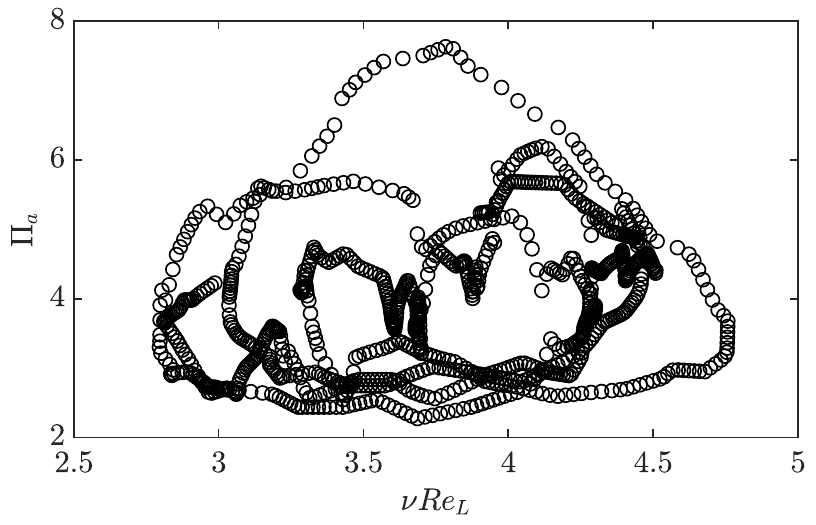}
		\end{tabular}
	}
	\caption{(a) Dissipation and (b) Large-scale flux, versus $Re_L$ for the forced turbulence simulation shown in figure \ref{fig:app2}b.}
	\label{fig:app3}
\end{figure*}

Additionally, data were retrieved from \cite{goto2016local}, for the case of forced periodic-box simulations. In that case, the flow quantities underwent quasi-periodic oscillations (see for instance figure \ref{fig:app2}b where $Re_\lambda$ and $C_\epsilon$ oscillate in anticorrelation, in accordance with the new dissipation scaling of equation \ref{eq:diss4}). In that case $\epsilon$ and $Re_L$ were found to be slightly anticorrelated (see figure \ref{fig:app3}a), whereas the large scale interscale flux $\Pi_a$ and $Re_L$ did not exhibit correlation (see figure \ref{fig:app3}b).  

Finally, data from the turbulence-grid experiments of \cite{valente2012universal} were retrieved and presented in figure \ref{fig:V1}c.

\section{Derivation of equation \ref{eq:L3}}
\label{app:3}

The classical dissipation scaling is $\epsilon = C_\epsilon u'^3/L$, with $C_\epsilon$ a constant. Combining this with assumption 2 ($\Pi_a = C_x u'^3/L$) one obtains $\Pi_a = C_x/C_\epsilon \epsilon$, with $C_x<C_\epsilon$. Similar to section \ref{sec:integral}, we differentiate equation \ref{eq:L1}, evaluate it at $kL=\kappa_a$ and inject it in the scale-by-scale energy budget (equation \ref{eq:budget}). Then, using the above relations for $\Pi_a$ and $\epsilon$ we obtain

$$
\frac{1}{\nu} \frac{dL^2}{dt} = -A' + B' Re^2 _\lambda \,,
$$

\noindent with $A' = \frac{4I_2}{I_0 \kappa_a f(\kappa_a,^*)}\frac{1-C_x/C_\epsilon}{\Phi}$ and $B'=\frac{2}{15}\frac{I_0C^2_\epsilon}{I_0-\kappa_af(\kappa_a,^*)}$. We inspect the sign of the denominator of $B'$. It can be checked from figure \ref{fig:A1}b, that $I_0>\kappa_af(\kappa_a,^*)$, or it can be approximately shown in the following way. Consider a model spectrum which, in the self-similar range ($\kappa_b>kL>\kappa_a$), assumes the form $f(\kappa,^*)=C\kappa^{-5/3}$, in accordance to  figure \ref{fig:A1}b. Then, $C = \kappa^{-5/3}_af(\kappa_a,^*)$. $I_0 = \int_{\kappa_a} ^{\kappa_b} f(\kappa,^*) d\kappa$ then becomes $I_0 = \frac{3}{2}\kappa_a f(\kappa_a,^*) \left[ 1- \left( \frac{\kappa_a}{\kappa_b} \right)^{2/3} \right]$, which is larger than $\kappa_a f(\kappa_a,^*)$ for small values of $\kappa_a/\kappa_b$. Thus, for sufficiently high Reynolds number, we expect $B'$ to be positive a constant.

\bibliography{Mybib}% Produces the bibliography via BibTeX.

\end{document}